\documentclass[aps,prx,twocolumn,10pt,showpacs,superscriptaddress,longbibliography]{revtex4-2}
\usepackage{hyperref}
\hypersetup{
    colorlinks = false,
    pdfborder = 0 0 0
}
\usepackage{amsmath}
\usepackage{amsfonts}
\usepackage{amssymb}
\usepackage{pifont}
\usepackage{mathtools}
\usepackage{bm}
\usepackage{cases}
\usepackage{braket}
\usepackage[version=4]{mhchem}
\usepackage{graphicx}
\allowdisplaybreaks%

\begin{document}
\title{Generic Spiral Spin Liquids}

\author{Xu-Ping Yao}
\affiliation{Department of Physics and HKU-UCAS Joint Institute 
for Theoretical and Computational Physics at Hong Kong, 
The University of Hong Kong, Hong Kong, China}
\author{Jian Qiao Liu}
\affiliation{Department of Physics and HKU-UCAS Joint Institute 
for Theoretical and Computational Physics at Hong Kong, 
The University of Hong Kong, Hong Kong, China}
\affiliation{International Center for Quantum Materials, 
School of Physics, Peking University, Beijing 100871, China}
\author{Chun-Jiong Huang}
\affiliation{Department of Physics and HKU-UCAS Joint Institute 
for Theoretical and Computational Physics at Hong Kong, 
The University of Hong Kong, Hong Kong, China}
\author{Xiaoqun Wang}
\affiliation{School of Physics and Astronomy, Tsung-Dao Lee Institute, 
Shanghai Jiao Tong University, Shanghai 200240, China}
\affiliation{Key Laboratory of Artificial Structures and Quantum Control of MOE,
Shenyang National Laboratory for Materials Science, Shenyang 110016, China} 
\author{Gang Chen}
\email{gangchen@hku.hk}
\affiliation{Department of Physics and HKU-UCAS Joint Institute for Theoretical 
and Computational Physics at Hong Kong, The University of Hong Kong, Hong Kong, China}
\affiliation{State Key Laboratory of Surface Physics and Department of Physics, 
Institute of Nanoelectronics and Quantum Computing, 
Fudan University, Shanghai, 200433, China}
 
\date{\today}

\begin{abstract}
    Spiral spin liquids are unique classical spin liquids that occur 
    in many frustrated spin systems, but do not comprise a new phase of matter. 
    Owing to extensive classical ground-state degeneracy, the
    spins in a spiral spin liquid thermally fluctuate cooperatively from a 
    collection of spiral configurations at low temperatures. These spiral 
    propagation wavevectors form a continuous manifold in reciprocal 
    space, \textit{i.e.}, a spiral contour or a spiral surface, that strongly governs
    the low-temperature thermal fluctuations and magnetic physics. 
    In this paper, the relevant spin models conveying the
    spiral spin liquid physics are systematically explored and the geometric origin of the spiral 
    manifold is clarified in the model construction. The spiral spin liquids 
    based on the dimension and the codimension of the spiral manifold are further clarified. 
    For each class, the physical properties are studied both generally 
    and for specific examples. The results are relevant to a wide range 
    of frustrated magnets. A survey of materials is given and future 
    experiments are suggested.
\end{abstract}

\maketitle

\section{Introduction}
\label{secI}

As an intriguing field in modern correlation physics, 
frustrated magnetism has attracted much interest 
in both theoretical and experimental efforts~\cite{Balents2010}.
In general, frustration describes various competing interactions 
in a system that cannot be satisfied simultaneously. 
For frustrated spin systems, these interactions refer to exchange 
interactions between the local magnetic moments.
It is expected that strong magnetic frustration could prevent the development 
of conventional magnetic orders and strongly suppress the ordering temperature. 
As a result, a wide range of exotic quantum phenomena may emerge,
including quantum phases such as quantum spin liquids~\cite{Savary_2016, Lee_2007, RevModPhys.89.025003} and the spin nematic phase~\cite{Kohama10686}. 
To understand these novel physics, new concepts represented by the internal 
gauge structure and fractionalized excitations have been introduced and developed 
in the past decades~\cite{Balents2010,Savary_2016,RevModPhys.89.025003}. 
Among these phenomena, quantum fluctuations play a significant role 
even in the absence of geometric frustration~\cite{KITAEV20062}.

\begin{figure}[b]
    \centering 
    \includegraphics[width=7.5cm]{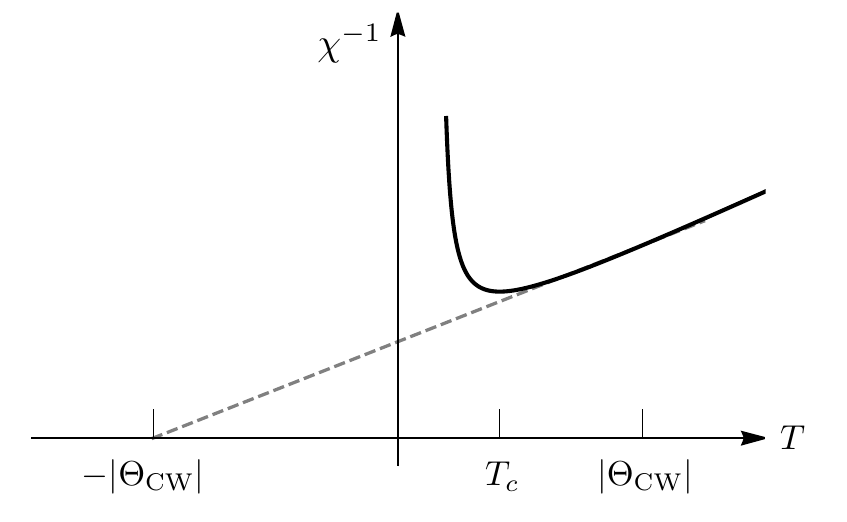}
    \caption{Magnetic susceptibility of a frustrated antiferromagnet. 
   The classical spin liquid (or cooperative paramagnet) 
   is located in the regime from $T_c$ to $\sim |\Theta_{\text{CW}}|$. 
   This would also be the regime for the spiral spin liquid
   if it is relevant for the underlying system. 
    }
    \label{susceptibility}
\end{figure}

\begin{table*}[t]
    \caption{Classification of spiral spin liquids. $d_s$ and $d_c$ refer to the
    dimension and codimension of the degenerate spiral manifold, respectively.
    ``TObD'' refers to the ``thermal order-by-disorder'' and $\Theta_{\text{CW}}$
    to the Curie-Weiss temperature.}
    \label{tab:summary}
    \begin{ruledtabular}
        \begin{tabular}{ccccccc}
        $d_s$ & $d_c$ & TObD & Stiffness & Low-temperature $C_v$ & Regime for SSL  & Model examples \\
        1 & 1 &  No & --- & $c_1+c_2 T$ & 0 $\sim  |\Theta_{\text{CW}}|$ & See Sec.~\ref{sec2A} and Sec.~\ref{sec4A} \\
        2 & 1 & Yes & $T^{2/3}$ & $c_1+c_2T^{1/3}$ & $T_\text{c}\sim |\Theta_{\text{CW}}|$ & Diamond lattice $J_1$--$J_2$ model~\cite{Bergman2007}  \\
        1 & 2 &  Yes & $T$ & $c_1+c_2T$ & $T_\text{c}\sim | \Theta_{\text{CW}}|$
        & See Sec.~\ref{sec2B}
        and Sec.~\ref{sec4A} \\
        \end{tabular}
    \end{ruledtabular}
\end{table*}

Surprisingly, when quantum fluctuation is not important or even absent, 
strong frustration still has the ability to realize rich and interesting physics. 
One such example is the spiral spin liquid that was first proposed for the 
$J_1$--$J_2$ Heisenberg model on the geometrically unfrustrated 
diamond lattice~\cite{Bergman2007}. In this spiral spin liquid, 
the classical ground states are coplanar spin spirals and have 
macroscopic degeneracy. The propagation vectors of the spiral spin 
configurations form a continuous manifold dubbed the spiral surface 
or spiral contour in reciprocal space. At finite temperatures, 
strong thermal fluctuations on the spiral manifold can stand out 
from the trivial paramagnetic regime. Hence, there could be  
a broad spiral spin liquid regime that smoothly connects to
the trivial paramagnet in the high-temperature regime, 
and this spiral spin liquid regime can manifest itself 
in the spin correlations and thermodynamic properties. 
Despite being quite interesting, the spiral spin liquid is in the thermal 
paramagnetic regime and is not really a new state of matter. 
In Fig.~\ref{susceptibility}, a representative behavior of the 
 magnetic susceptibility of a typical
frustrated magnet is depicted in which the spiral spin liquid would be located in the regime
from the ordering temperature $~T_c$ 
to the Curie-Weiss temperature $|\Theta_{\text{CW}}|$. 
At very low temperatures, the spiral orderings may develop from so-called 
thermal order by disorder or from other sub-leading interactions~\cite{Bergman2007,PhysRevB.78.144417}. 
The external magnetic field could further select a combination of 
equivalent and coplanar magnetic wavevectors and stabilize 
non-trivial topological spin structures such as vortexes, 
merons, and skyrmion lattices~\cite{PhysRevB.93.085132, PhysRevB.100.224404, Gao2016, PhysRevLett.123.057202,Gao_2020}.

Recently, the degenerate spiral surface has been confirmed directly 
by neutron-scattering experiments in a long-promised spin-5/2 
diamond-lattice compound 
\ce{MnSc2S4} \cite{Gao2016,Gao_2020} with the single-crystal sample 
 nine years after the theoretical proposal~\cite{Bergman2007}. 
 This caused a new wave of attention to be paid to the spiral spin liquids, the proximate orders, 
 and the underlying novel physics~\cite{PhysRevLett.122.097201,bordelon2020frustrated, 
 PhysRevLett.123.057202, PhysRevB.97.115102, pohle2017spin,PhysRevB.81.214419,
 PhysRevB.100.140402,PhysRevResearch.2.043278,Niggemann_2019,Nussinov}. 
 Although many relevant models and materials have been well studied in past decades, 
a systematic study for the flourishing family of the spiral spin liquids 
is still lacking. A recent theoretical work by Attig \textit{et al.} 
gives a quite complete list of the spin models conveying  
the spiral manifold~\cite{PhysRevB.96.085145}. What the authors 
of Ref.~\onlinecite{PhysRevB.96.085145} focus on are the 
spin-fermion correspondence and patterns of spiral 
manifolds at zero temperature. The resulting behaviors 
at finite temperatures still remain to be understood. The purpose 
of this work is to classify the spiral spin liquids, as well as to illustrate 
the characteristics of their finite-temperature behaviors by constructing 
the minimal models. The basic idea is to consider the dimensions 
and codimensions of the spiral manifold (see Table.~\ref{tab:summary}). 
The spin interactions are then designed accordingly and the massively degenerate ground-state manifold effectively manipulated. 
For both spiral contour and spiral surface cases, 
a large regime is identified in which spiral spin liquid physics 
dominates up to a high temperature compared to interactions.
The robustness of the spiral spin liquid features is emphasized, 
which is important for the realization and identification in the
realistic materials. Also discussed are the thermal order-by-disorder 
mechanism~\cite{Villain1980} and its consequences in 
three-dimensional systems. It is shown that 
thermodynamic quantities such as the specific heat would 
be modified by the induced spiral ordering, and that a distinct 
temperature dependence is exhibited for spiral manifolds with different dimensions.
The scenario of the energy-entropy competition might be more intricate 
than as demonstrated herein, but is believed to be ubiquitous in real 
materials~\cite{Villain1980, PhysRevLett.62.2056, PhysRevB.48.9539}.

The rest of this paper is organized as follows. The spiral spin liquids are reviewed and classified in Sec.~\ref{secII}. 
For each class, a general Heisenberg Hamiltonian is constructed
to capture the essential physics of a spiral manifold at zero temperature. 
As quintessences, the square, honeycomb, and AB-stacked multilayered triangular lattices are analyzed in detail. In Sec.~\ref{secIII}, 
phase diagrams with the temperature and identity of
the spiral spin liquid regimes for each case are further established. 
The difference in the low-temperature specific heat is discussed.
Finally, in Sec.~\ref{secIV}, this work concludes with a survey
of the existing and potential materials with spiral spin liquid physics, and a discussion about open theoretical questions.

\section{Spiral spin liquids: Classification and model construction}
\label{secII}

The basic structure underlying spiral spin liquids is the 
continuous spiral manifold formed by the propagation wavevectors 
of the classical ground states in reciprocal space. 
The continuity of this spiral manifold reflects the massive 
ground-state degeneracy and promises a spiral spin liquid. 
As will be shown later, the relationship between the dimensions and 
codimensions of the spiral manifold is crucial for the physical 
properties of the systems. The codimension $d_c$ 
of the spiral manifold is defined as the difference between the dimension of 
the system and the dimension ($d_s$) of the spiral manifolds, and 
the spiral spin liquids are classified based on $(d_s,d_c)$. 
For example, the well-known $J_1$--$J_2$ model 
on the diamond lattice for \ce{MnSc2S4} has 
a degenerate spiral surface with a dimension of $d_s=2$, 
and the codimension of this spiral surface is $d_c=1$. 
This physical classification is listed in Table.~\ref{tab:summary}. 
In this section, specific models are proposed to explicitly demonstrate 
the existence of other classes and study their physical properties. 
The square, honeycomb, and AB-stacked multilayered 
triangular lattices are analyzed in detail.

\subsection{Class with $d_s=1$ and $d_c=1$}
\label{sec2A}

% 2d system with line degeneracy

First, a generic $J_1$--$J_2$ 
Heisenberg model on the bipartite lattices with the following form is considered: 
\begin{equation}
\label{HamA}
    H=\frac{1}{2}\sum_{i,j}\left[ J_1 \bm{M}_{ij} + J_2 (\bm{M}^2 
    - z \mathbb{I})_{ij}\right]\bm{S}_i\cdot\bm{S}_j, 
\end{equation}
where $\bm{S}_i$ is the classical three-component spin vector with unit length. 
$\bm{M}$ is a symmetric adjacency matrix of the underlying lattice, defined as
\begin{equation}
	\bm{M}_{ij}=\bm{M}_{ji}=
	\begin{cases}
	1 & \text{for } \braket{ij}, \\
	0 & \text{otherwise}, 
	\end{cases}
\end{equation}
where $\braket{ij}$ refers to the bond connecting the nearest-neighbor sites 
$i$ and $j$, $\mathbb{I}$ represents the identity matrix, and $z$ is the 
coordination number of the lattice. The geometric meaning of 
${\bm{M}^2-z\mathbb{I}}$ is clear.
Since ${(\bm{M}^2)_{ij}=\sum_{k}\bm{M}_{ik}\bm{M}_{kj}}$, 
$(\bm{M}^2)_{ij}$ essentially counts the number of different paths 
hopping from site $i$ to site $j$ via two nearest-neighbor bonds. 
When $(\bm{M}^2)_{ii} = z\mathbb{I}$, this represents the processes 
hopping back to the starting site. Therefore, these returning processes are subtracted
from $\bm{M}^2$. Hereafter, it is assumed that the antiferromagnetic interactions are $J_1>0$ 
and $J_2>0$. 
The regime with $J_1<0$ and $J_2>0$ can be achieved by only flipping spins 
in one sublattice, and there is no frustration when $J_2<0$.
It will be seen that this model encodes the essential ingredients 
of the spiral manifold, and can be constructed on different lattices.

\begin{figure}[t]
    \centering 
    \includegraphics[width=0.45\textwidth]{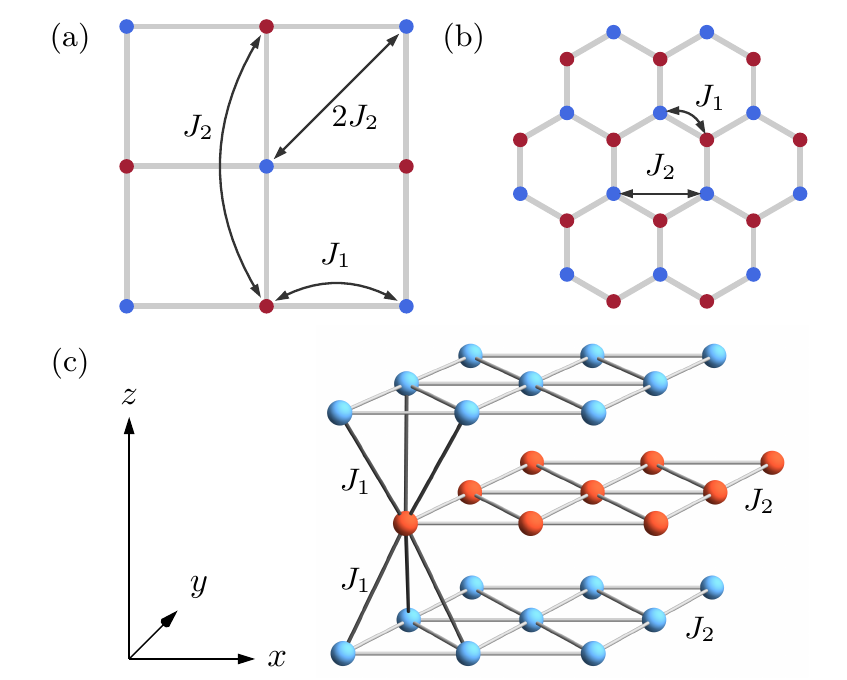}
    \caption{(Color online.) (a) Square, (b) honeycomb, and (c) AB-stacked multilayered triangular lattices with $J_1$ and $J_2$ spin-exchange interactions. Sublattices are colored red and blue. Only interlayer bonds for one site are plotted in (c) for clarity. Lattice geomatries and crystal momenta are defined in Appendix~\ref{appI}.}
    \label{fig:lattices}
\end{figure}

The well-known Luttinger-Tisza method is employed to explore the ground 
states of the classical $J_1$--$J_2$ Heisenberg model in Eq.~\eqref{HamA}. 
The core of this method is softening the local unit-length constraint 
on classical spin vectors ${\bm{S}_i^2=1}$ by a global constraint 
${\sum_i |\bm{S}_i|^2=N}$, where $N$ is the total number of spins. 
After this replacement, the original problem is reduced to minimizing the energy
\begin{equation}
    E_{\text{LT}}=H-\lambda\left(\sum_i |\bm{S}_i|^2-N\right),
\end{equation}
where $\lambda$ is a Lagrange multiplier. 
Fourier transformation is implemented on each spin vector with the sublattice index $\mu$ and unit-cell index $i$ as 
\begin{equation}
    \bm{S}_{i}^{\mu}=\frac{1}{\sqrt{N/2}}\sum_{\bm{k}}\bm{S}^{\mu}_{\bm{k}}e^{i\bm{k}\cdot\bm{r}_i}.
\end{equation}
Then, the energy in reciprocal space reads
\begin{equation}
    E_{\text{LT}}=\sum_{\bm{k}}\left[\frac{1}{2}\mathcal{J}_{\mu\nu}(\bm{k})-\lambda\delta_{\mu\nu}\right]\bm{S}_{\bm{k}}^{\mu}\bm{S}_{-\bm{k}}^{\nu}+\lambda N,
\end{equation}
where $\mathcal{J}_{\mu\nu}(\bm{k})$ is the Fourier transformation of the interaction matrix defined by the adjacency matrix $\bm{M}$ in real space.
For the model proposed here, it has the simple form 
\begin{equation*}
    \mathcal{J}(\bm{k})=
    \begin{bmatrix}
        J_2 \big(\Lambda(\bm{k})^2-z \big) & J_1 \Lambda(\bm{k})e^{i\theta(\bm{k})} \\
        J_1 \Lambda(\bm{k})e^{-i\theta(\bm{k})}  & J_2 \big(\Lambda(\bm{k})^2-z \big)
    \end{bmatrix},
\end{equation*}
where the parameter 
${\Lambda(\bm{k})=\left|\sum_{\bm{R}_{\text{NN}}}e^{-i\bm{k}\cdot\bm{R}_{\text{NN}}}\right|}$. 
$\bm{R}_{\text{NN}}$ denotes the nearest-neighbor vectors 
and $\theta(\bm{k})$ refers to the phase shift between two sublattices.
Now, minimizing the energy $E_{\text{LT}}$ is equivalent to finding the minimum 
eigenvalue of a quadratic form $\mathcal{J}(\bm{k})$.
It is found that the eigenvalues form two bands with dispersions as
\begin{equation}
    \omega_{\pm}=J_2\left[\Lambda(\bm{k})\pm\frac{J_1}{2J_2}\right]^2-\frac{J_1^2+4zJ_2^2}{4J_2}.
\end{equation}
A phase transition occurs at ${J_2/J_1=1/(2z)}$. 
When ${J_2/J_1<1/(2z)}$, $\omega_{-}$ takes its minimum 
at the $\Gamma$ point in reciprocal space, corresponding 
to the antiferromagnetic Ne\'{e}l ordered state for the bipartite lattices. 
With increasing $J_2/J_1$, a massive ground-state degeneracy 
emerges and manifests itself as a spiral manifold determined by
\begin{equation}
    \Lambda(\bm{k})=\frac{J_1}{2J_2}.
\end{equation}

The spin configuration is then decoded from the corresponding eigenvectors. 
Given a wavevector $\bm{Q}$ that resides on the degenerate manifold, 
the ground-state spin configuration in real space takes a spiral form,
\begin{multline}
    \bm{S}_i^{\mu}=(-1)^{\mu}
    \cos\left[\bm{Q}\cdot\bm{r}_i+(-1)^{\mu}\theta(\bm{Q}/2)\right]\bm{e}_1 \\
    +(-1)^{\mu}\sin\left[\bm{Q}\cdot\bm{r}_i+(-1)^{\mu}\theta(\bm{Q}/2)\right]\bm{e}_2,
\end{multline}
where $\mu$ can take $0$ and $1$ for two sublattices, respectively. 
$\bm{e}_{1,2}$ are any two orthogonal unit vectors. 

In this subsection, the class with $d_s=1$ and $d_c=1$ is studied, where the spiral manifold
is a contour in reciprocal space. 
In this class, the dimension of the system is 2. 
To achieve this, only one constraint on the spiral wavevector
 is needed to force the spins
onto the spiral manifold for the classical ground states. 
The square and honeycomb lattices are explicitly presented
as examples that are shown in Figs.~\ref{fig:lattices}(a) and (b). 
The square and honeycomb lattices are both bipartite and 
have the coordination numbers ${z=4}$ and ${z=3}$, respectively. 
The nearest-neighbor interaction $J_1$ and 
the next-nearest-neighbor interaction $J_2$ are considered for both lattices.
In addition, a third nearest-neighbor interaction 
with the same strength of $J_2$ for the square lattice is involved to 
satisfy the definition of the adjacency matrix on the square lattice. 
The resulting spiral contour parameters are
\begin{equation}
    \Lambda(\bm{k})=4 \big|\cos(\bm{a}_1\cdot 
    \frac{\bm{k}}{2})\cos(\bm{a}_2\cdot \frac{\bm{k}}{2}) \big|
\end{equation}
for the square lattice and
\begin{multline}
    \Lambda(\bm{k})= \big[3+2\cos(\bm{a}_1\cdot\bm{k})+2\cos(\bm{a}_2\cdot\bm{k}) 
     \\
    +2\cos[(\bm{a}_1-\bm{a}_2)\cdot\bm{k}]\big]^{\frac{1}{2}}
\end{multline}
for the honeycomb lattice. 
Here, $\bm{a}_{1}$ and $\bm{a}_2$ 
are the lattice vectors defined in Appendix.~\ref{appI}. 
In Figs.~\ref{fig:square}(a) and~\ref{fig:honeycomb}(a), 
the spiral contours for the two lattices and their 
evolution with increasing $J_2/J_1$ are plotted. 
It is shown that both two spiral contours emerge at ${J_2/J_1=1/(2z)}$ 
from the $\Gamma$ point and expand towards the boundary of 
the respective Brillouin zones. Despite this similarity, the 
spiral contours of two lattices behave dramatically differently 
in the large-$J_2/J_1$ regime. For the square lattice, 
the spiral contour approaches the boundary continuously, 
but does not touch it until the ${J_2/J_1\rightarrow \infty}$ limit, where
the spiral contour completely coincides with the Brillouin-zone boundary. 
Physically, in this limit the proposed generic $J_1$--$J_2$ Heisenberg model on 
the square lattice reduces to a conventional one with the largest frustration, 
and the massive degeneracy of the ground states remains. 
On the contrary, the spiral contour of the honeycomb lattice~\cite{PhysRevB.81.214419} touches 
the $M$ point and other symmetry-related points in reciprocal 
space at ${J_2/J_1=1/2}$, and then splits into six arcs. 
These arcs are symmetry equivalent and can be viewed as closed 
loops centered at the $K$ points. They shrink with increasing $J_2/J_1$ and finally become isolated points residing at point $K$. 
This means that the ground-state degeneracy disappears and N\'{e}el 
ordering is restored. Actually, in the ${J_2/J_1\rightarrow \infty}$ limit, 
the two sublattices of the honeycomb lattice can be decoupled into 
two antiferromagnets on the triangular lattice, which is consistent with 
the present results. Therefore, the most important difference of these two 
types of lattices is whether the largest frustration state can be achieved 
at finite $J_2/J_1$, where the spiral contour has the largest circumference.

\subsection{Class with ${d_s=1}$ and ${d_c=2}$}
\label{sec2B}

\begin{figure}[t]
\centering 
\includegraphics[width=0.45\textwidth]{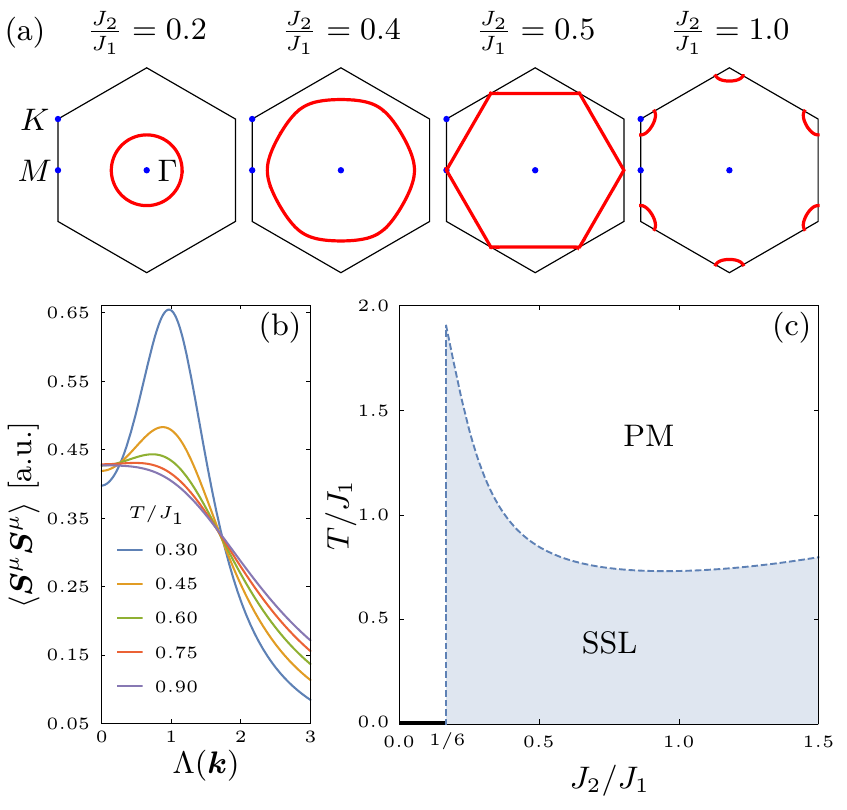}
    \caption{(Color online.) Honeycomb lattice. (a) Evolution of spiral contour (red) with couplings in the first Brillouin zone. High-symmetry $\bm{k}$ points are marked as blue dots. (b) Spin correlation as a function of $\Lambda(\bm{k})$ at different temperatures. Here, $J_2/J_1=0.5$ is fixed where the spin correlation is peaked at $\Lambda (\bm{k}) =1$.  
(c) Phase diagram with temperature. Spiral spin liquid (SSL) regime is 
    outlined in the paramagnetic phase (PM). Black line refers to N\'{e}el ordered state.}
    \label{fig:honeycomb}
\end{figure}

The class with ${d_s=1}$ and ${d_c=2}$ leads to a spiral contour 
on a three-dimensional lattice. Such a model can be constructed 
from a two-dimensional bipartite lattice. Specifically, 
the honeycomb lattice is decoupled into two triangular sub-systems 
according to its sublattices. The sublattice layers are then stacked
alternatively into a three-dimensional multilayered system, as    
shown in Fig.~\ref{fig:lattices}(c). This is known as AB stacking. 
Here, the projection of a site from one layer onto its adjacent layer 
is located at the center of the triangles. The spin Hamiltonian retains 
the $J_1$--$J_2$ Heisenberg form
\begin{equation}
\label{HamB}
    H=J_1\sum_{i,j}\bm{S}_i\cdot\bm{S}_j+J_2\sum_{i,j}\bm{S}_i\cdot\bm{S}_j,
\end{equation}
where $J_1>0$ ($J_2>0$) then becomes the interlayer (intralayer) interaction.
Repeating the process of the Luttinger-Tisza method, 
it is found that the interaction matrix is
\begin{equation}
    \mathcal{J}(\bm{k})=
    \begin{bmatrix}
        J_2\big(\frac{1}{4}\Lambda(\bm{k}_{\parallel})^2-3\big) & J_1^{\prime}\Lambda(\bm{k}_{\parallel})e^{i\theta(\bm{k})} \\
        J_1^{\prime}\Lambda(\bm{k}_{\parallel})e^{-i\theta(\bm{k})} & J_2\big(\frac{1}{4}\Lambda(\bm{k}_{\parallel})^2-3\big)
    \end{bmatrix},
\end{equation}
where ${\bm{k}_{\parallel}=\{k_x,k_y,0\}}$ and 
$J_1^{\prime}=J_1 \cos\big(\frac{1}{2}\bm{a}_3\cdot\bm{k}\big)$ 
with the lattice vectors $\bm{a}_i$ defined in Appendix~\ref{appI}. 
The explicit form of $\Lambda(\bm{k}_{\parallel})$ is expressed as
\begin{multline}
    \Lambda(\bm{k})=2 \left[ 3+2\left[\cos(\bm{a}_1\cdot\bm{k}_{\parallel})+\cos(\bm{a}_2\cdot\bm{k}_{\parallel})\right]\right. \\
    \left. +\cos\left[(\bm{a}_1-\bm{a}_2)\cdot\bm{k}_{\parallel}\right]\right]^{\frac{1}{2}}.
\end{multline}
The minimum eigenvalue of $\mathcal{J}(\bm{k})$,
\begin{equation}
    \omega_{-}=\frac{J_2}{4}\left[\Lambda(\bm{k}_{\parallel})-\frac{2J_1^{\prime}}{J_2}\right]^2-\frac{J_1^{\prime 2}+3J_2^2}{J_2},
\end{equation}
defines a spiral contour
\begin{equation}
\Lambda({\bm{k}_\parallel})=\frac{2J_1}{J_2} ,
\end{equation}
on the ${k_z=0}$ plane when ${J_2/J_1>1/3}$. 
Before the spiral contour emerges, an antiferromagnetic 
N\'{e}el ordering is favored. As depicted in Fig.~\ref{fig:ABcontour}, 
the evolution of the spiral contour is very similar to that of the honeycomb 
lattice shown in Fig.~\ref{fig:honeycomb}(a). 
Nevertheless, the critical value at which the spiral contour touches
the $M$ points and is reconstructed becomes ${J_2/J_1=1}$. 
This can be attributed to the layer structure with doubling interlayer 
(inter-sublattice) bonds compared to the honeycomb lattice.

\subsection{Examples with $d_c=0$}
\label{sec2C}

In addition to the classes listed in Table.~\ref{tab:summary}, 
there exist examples with ${d_c=0}$. The well-known cases 
are the classical Kagom\'{e} lattice Heisenberg model and 
classical pyrochlore lattice Heisenberg model. Because of the 
local zero-energy modes, the classical ground state has
extensive degeneracy. The lowest eigenvalue of the 
exchange matrix in reciprocal space has no dispersion. 
The flatness arises from the existence of local zero-energy modes
on each hexagon plaquette in the ground-state manifold. 
It is better to undersand the classical ground state in the 
real-space picture. The classical ground state requires that
each triangular plaquette of the Kagom\'{e} lattice 
has a total spin equal to zero. This is like a Gauss's law constraint
for electromagnetism. In fact, the low-temperature magnetic
properties can be well captured by an emergent classical electromagnetism
that gives rise to a power-law spin correlation with a dipolar-like 
angular dependence. The classical pyrochlore lattice Heisenberg model
was understood in a similar fashion. 
For the above reason, these two ${d_c=0}$ 
examples are not included in the current study of spiral spin liquids.

\begin{figure}[t]
    \centering 
    \includegraphics[width=0.45\textwidth]{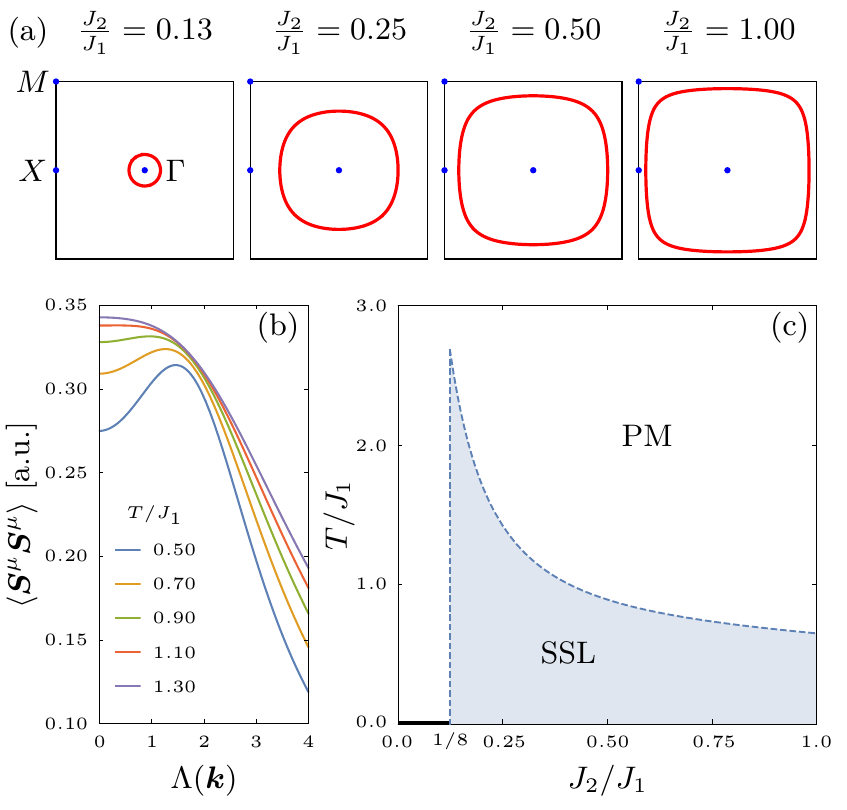}
    \caption{(Color online.) Square lattice. (a) Evolution of spiral contour (red) with couplings in the first Brillouin zone. High-symmetry $\bm{k}$ points are marked as blue dots. (b) Spin correlation as a function of $\Lambda(\bm{k})$ at different temperatures. Here, $J_2/J_1=0.3$ is fixed and the spin correlation is peaked at ${\Lambda (\bm{k}) = 5/3}$. (c) Phase diagram with temperature. Spiral spin liquid (SSL) regime is outlined in the paramagnetic phase (PM). Black line refers to N\'{e}el ordered state.}
    \label{fig:square}
\end{figure}

\section{Finite-temperature behaviors}
\label{secIII}

In the preceding section, the similarity and difference 
between different classes of spiral spin liquids were discussed. 
Next, the finite-temperature behaviors of the spiral spin liquid regime are explored. 
A crucial difference here is the existence of ordered states 
due to the thermal order-by-disorder in the three spatial dimensions, 
which is prohibited in two spatial dimensions by the well-known 
Mermin–Wagner theorem. These ordered states can 
only persist in the rather low-temperature regime in three dimensions. 
In general, the increasing thermal fluctuation will eventually favor a disordered 
state as temperature increases. 
It will be shown that a broad range of the spiral spin liquid regime 
can persist before crossing over to the trivial paramagnetic regime.

\subsection{Spiral spin liquid regime}
\label{sec3A}

\begin{figure*}[t]
    \includegraphics[width=\textwidth]{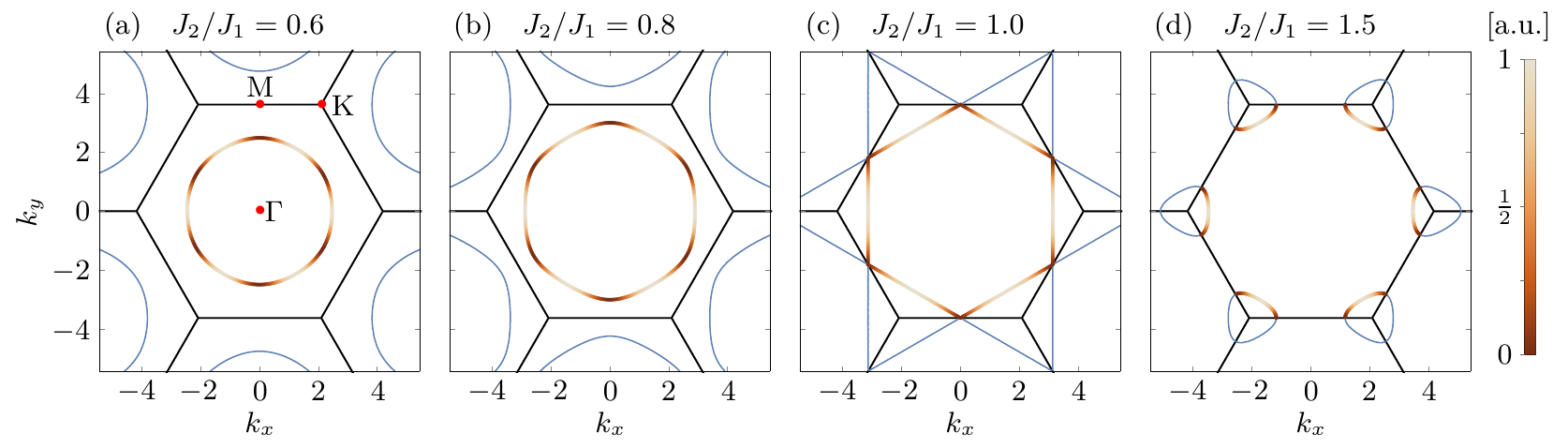}
    \caption{(Color online.) Spiral contour for AB-stacked multilayered triangular lattice with different $J_2/J_1$ on the ${k_z=0}$ plane. 
    Black lines refer to the Brillouin-zone boundaries. 
    High-symmetry $\bm{k}$ points 
    are marked as red dots. In the first Brillouin zone,
     the spiral contours are colored according to the relative magnitudes of
     the free energy.}
    \label{fig:ABcontour}
\end{figure*}

The presence of the massive ground-state degeneracy, \textit{i.e.}, 
the spiral surface manifold, will affect the finite-temperature behavior 
of the system by revealing itself in different ways in different temperature regimes. 
For two spatial dimensions, the Mermin-Wagner theorem implies the absence of 
long-range order at finite temperatures, and there should be a wide regime 
in which the spiral spin liquid dominates the physics down to zero temperature 
when $J_2/J_1>(J_2/J_1)_{\text{c}}$. The stability of the spiral spin liquid regime 
can be extended to the three-dimensional case, except that
ordered states may exist at a sufficiently low temperature, which 
will be explained later. Here, the spiral spin liquid regime 
is characterized by the strong peaks of the spin-spin correlation 
in reciprocal space, located exactly on the spiral manifold. 
As the thermal fluctuation increases, these peaks will gradually 
become flat and broad, and will eventually become invisible 
in the trivial paramagnetic limit. To elucidate the regime 
of the spiral spin liquid, the spin correlation function 
at high temperatures is calculated by the self-consistent Gaussian approximation. 
Under the global constraint as introduced in Sec.~\ref{secII}, 
the partition function of the spin system is given by 
\begin{equation}
    \mathbb{Z}=\int\mathcal{D}S  \mathcal{D}\lambda 
    \, 
    e^{-\beta H-i\lambda(\sum_i S_i^2-N)}
\end{equation}
where $\lambda$ is the Lagrange multiplier to enforce 
${\sum_i {\boldsymbol S}_i^2 = N}$. This is also known 
as the spherical approximation~\cite{Bergman2007}.
The saddle-point solution can be obtained by integrating out the 
classical spins, and the effective action scales as $N$ that naturally
leads to the saddle-point solution for $\lambda$.
The saddle-point solution is taken by replacing 
$i\lambda\rightarrow\beta\lambda'(T)/2$, where $\lambda'(T)$ 
is a saddle-point parameter to be determined self-consistently. 
Equivalently, the saddle-point equation can be obtained from the 
spherical constraint for the spins,
\begin{align}
    \sum_i \bm{S}_i^2= & \sum_{\mu}\sum_{\bm{k}}S_{\bm{k}}^{\mu}S_{-\bm{k}}^{\mu} \notag \\
    = & \sum_{\bm{k}}\frac{3}{\beta}\left[\frac{1}{\omega_{-}(\bm{k})+\lambda'(T)}+\frac{1}{\omega_{+}(\bm{k})+\lambda'(T)}\right] \notag \\
    \equiv & \, N,
    \label{eq17}
\end{align}
where the spin-correlation function in 
reciprocal space for the spins from the same sublattice $\mu$ is  
\begin{equation}
    \braket{\bm{S}^{\mu}_{\bm{k}}\bm{S}^{\mu}_{-\bm{k}}}=\frac{1}{2\beta}\left[\frac{1}{\omega_{-}(\bm{k})+\lambda'(T)}+\frac{1}{\omega_{+}(\bm{k})+\lambda'(T)}\right],
\end{equation}
and $\omega_{\pm}$ refers to the two branches of eigenvalue dispersions
of the exchange-interaction matrix $\mathcal{J}(\bm{k})$ defined in 
Sec.~\ref{secII}. 
One can obtain the parameter $\lambda'(T)$ by solving the 
above saddle-point equation. 
In Figs.~\ref{fig:square}(b) and~\ref{fig:honeycomb}(b),
the temperature dependence of the spin correlation $\braket{\bm{S}^{\mu}_{\bm{k}}\bm{S}^{\mu}_{-\bm{k}}}$ is depicted as a function of $\Lambda(\bm{k})$ for the square and honeycomb lattices.  The peaks correspond to points near the spiral contour that 
are apparent and remain discernible up to a crossover temperature.
The spiral spin liquid regime can then be roughly identified by tracing this feature. 
The crossover between the spiral spin liquid regime and 
conventional high-temperature paramagnet regime is outlined in 
Figs.~\ref{fig:square}(c) and~\ref{fig:honeycomb}(c). 
It is shown that the frustration from the competing spin interactions 
can manifest itself qualitatively through this crossover temperature. 
The enhancement of frustration will strongly suppress 
the spiral spin liquid physics and make it easier for the system    
to enter a trivial paramagnetic state under the thermal fluctuation. 
In proximity to ordered states, a spiral spin liquid can persist up to 
a higher temperature. Since the largest frustration for the square 
lattice appears in the ${J_2/J_1\rightarrow\infty}$ limit, 
a monotonically decreasing crossover temperature is obtained, 
in contrast to the honeycomb-lattice case in which 
there is a minimal crossover temperature near ${J_2/J_1=1/2}$.
The spiral spin liquid regime is a special paramagnetic phase in which 
the spiral manifold governs the thermal fluctuations.

The finite-temperature behaviors for three dimensions have no qualitative difference. 
For the proposed AB-stacked multilayered triangular lattice, the spiral spin liquid regime and its crossover
to the trivial paramagnet resemble those of the honeycomb lattice. The difference is
that, in the low-temperature regime, the spiral ordered states can emerge through the thermal order-by-disorder mechanism. This is discussed in the next subsection.

\subsection{Thermal order-by-disorder}
\label{sec3B}

Although the ground-state configuration can fluctuate on the degenerate spiral manifold without any energy cost at zero temperature, a global selection among these ground states will occur when weak thermal fluctuation is considered. 
At finite temperatures, the free energies of the degenerate ground states are usually different. 
To minimize the free energy, certain spiral ordered states associated with large entropy are selected. 
This stabilization mechanism of long-range order driven by the entropy is termed thermal order-by-disorder.
This effect for the AB-stacked multilayered triangular lattice model is studied in the low-temperature regime.

First, a spiral spin order at momentum $\bm{q}$ with a spin configuration $\bar{\bm{S}}$ lying in the $x$--$y$ plane is assumed and then its local stability is analyzed. The fluctuations can be parameterized as
\begin{equation}
    \bm{\phi}=\phi^{\text{o}}\hat{\bm{z}}+\phi^{\text{i}}[\hat{\bm{z}}\times\bar{\bm{S}}],
\end{equation}
where $\phi^{\text{o}}$ and $\phi^{\text{i}}$ are out-of-plane and in-plane fluctuations, respectively. 
Physical spins can be expressed in fluctuation variables around the assumed order as 
\begin{equation}
    \bm{S}=\bm{\phi}+\bar{\bm{S}}(1-\bm{\phi}^2)^{1/2}.
\end{equation}
The unit-vector constraint remains satisfied since $\bm{\phi}\cdot\bar{\bm{S}}=0$. For $\bm{\phi}^2\ll 1$, one can safely expand the spin Hamiltonian~\eqref{HamB} in a fluctuation field up to second order,
\begin{equation}
    H= \frac{1}{2}\sum_{ij}\tilde{J}_{ij}\phi^{\text{o}}_i\phi^{\text{o}}_j+\tilde{J}_{ij}(\bar{\bm{S}}_i\cdot\bar{\bm{S}}_j)\phi^{\text{i}}_i\phi^{\text{i}}_j.
\end{equation}
Here, $\tilde{J}_{ij}=J_{ij}-\delta_{ij}\omega_{-}^{\text{min}}$. The shift makes both eigenvalues non-negative and only scales the partition function by a constant. 
There is no change in free energy, except for a zero-point movement.
The zeroth- and first-order Hamiltonians have also been ignored because the former is just the ground-state energy and the latter should vanish in an ordered state.
The resulting partition function is
\begin{equation}
    \mathbb{Z}=\int\mathcal{D}\bm{\phi}e^{-\beta H}(1-\bm{\phi}^2)^{-1/2}\propto\int\mathcal{D}\phi^{\text{o}}\mathcal{D}\phi^{\text{i}} e^{-\mathcal{S}},
\end{equation}
where the action is $\mathcal{S}=\beta H$. 
In the last step, a Jacobian factor is dropped since $(1-\bm{\phi}^2)^{-1/2}\approx e^{(\phi^{\text{o}2}+\phi^{\text{i}2})/2}$ up to second order. 
That means that this factor only contributes diagonal terms to the action and does not affect the relative magnitude of free energy.
At the Gaussian level, two components of the fluctuation field are decoupled. 
The out-of-plane fluctuation modes $\phi^{\text{o}}$ have zero modes for all momenta on the spiral manifold and thus are responsible for the massive ground-state degeneracy. 
The in-plane fluctuation modes $\phi^{\text{i}}$, however, have only one single gapless mode and are well-behaved at low temperature. 
The gapless mode is just the Goldstone mode required by symmetries and indicates a long-range order. 

For the proposed $J_1$--$J_2$ Heisenberg model on the AB-stacked multilayered triangular lattice, these $\phi^{\text{i}}$ fluctuation modes with spiral ordering wavevector $\bm{q}$ are 
\begin{multline}
    \gamma_{\pm}(\bm{k},\bm{q})=\frac{J_2}{8}\left(|\xi_{+}|^2+|\xi_{-}|^2-24\right)-\omega_{-}^{\text{min}} \\
    \pm \frac{J_1}{2}\left|\xi_{+}\cos\frac{\bm{a}_3\cdot(\bm{k}+\bm{q})}{2}+\xi_{-}\cos\frac{\bm{a}_3\cdot(\bm{k}-\bm{q})}{2}\right|,
\end{multline}
where $\xi_{\pm}=\Lambda(\bm{k}_{\parallel}\pm\bm{q})e^{\mp i\theta(\bm{q})+i\theta(\bm{k}\pm\bm{q})}$. Therefore, one can integrate the fluctuation fields to obtain the $\bm{q}$-dependent part of free energy,
\begin{equation}
    F_{\phi^{\text{i}}}(\bm{q})=T\int_{\bm{k}}[\ln\gamma_{+}(\bm{k},\bm{q})+\ln \gamma_{-}(\bm{k},\bm{q})]/(2\pi T).
\end{equation} 
The free-energy distribution on the spiral contour is visualized as shown in Fig.~\ref{fig:ABcontour}, in which the color reflects the relative magnitude of $F_{\phi^{\text{i}}}(\bm{q})$. 
Once the spiral contour emerges, the free-energy minima arise at $(0,q,0)$ and symmetry-related points, indicating a phase transition from N\'{e}el order to spiral order. 
When the spiral contour touches the Brillouin-zone boundary at $J_2=J_1$, the minima locate at $M$ points and move towards $K$ points as $J_2/J_1$ increases. 
In the $J_2/J_1\rightarrow \infty$ limit, two decoupled antiferromagnetic magnets on triangular lattices are obtained, as expected.

\begin{figure}[t]
    \centering
    \includegraphics[width=0.45\textwidth]{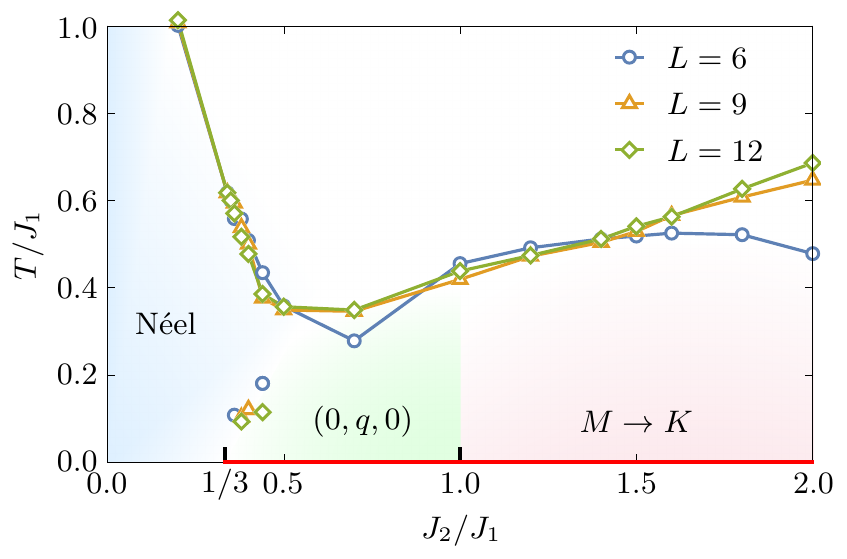}
    \caption{(Color online.) Phase diagram for AB-stacked multilayered triangular lattice in low-temperature regime. In addition to N\'{e}el order, two types of spiral order are identified in agreement with the current thermal order-by-disorder analysis. The ordering temperatures $T_c$ are determined to be locations of specific-heat peaks by classical Monte Carlo simulations for systems with $L=6,9,12$. Near the critical value of $J_2/J_1=1/3$, a re-entrant of N\'{e}el order occurs. Red line refers to spiral spin liquid phase at zero temperature.}
    \label{fig:stack}
\end{figure}

The spiral ordering selected by entropy will melt when the system leaves the low-temperature regime. 
Above a critical temperature $T_c$, the spiral spin liquid behavior prevails again. 
%To construct a phase diagram for spin orderings, we carry out classical Monte Carlo simulations on a layer-stacked triangular lattice with $L^3$ unit cells. 
%The numeric results are shown in Fig.~\ref{fig:stack}.
To construct a finite-temperature phase diagram for spin orderings, classical Monte Carlo simulations are carried out on the AB-stacked multilayered triangular lattice with $L^3$ unit cells. 
Owing to strong frustrations, a simple Metropolis-Hastings update will have a very long auto-correlation time that would lead to non-ergodicity.  
Here, a generally applicable method, parallel tempering~\cite{Koji1996}, 
is employed to resolve this issue. Essentially, in parallel tempering 
the simulated space consists of several replicas of the systems 
while they are at different temperatures. 
Exchanging configurations in different replicas will decrease or increase the total energy according to the detailed balance, and there is an acceptance probability for this update. 
Therefore, the configurations of the high-temperature replicas can be tunneling to 
the lower-temperature replicas and vice versa. 
This diffusion process will significantly improve the efficiency. 
Furthermore, the over-relaxation method is employed to further decrease 
the auto-correlation time as well.
The finite-temperature phase boundary can be determined by the grown peak of 
the specific heat with increasing system size, and the numerical results are shown in 
Fig.~\ref{fig:stack}. Compared to the rapidly declining boundary of N\'{e}el order, 
$T_c$ has a slower growth in the spiral order phases.
The lowest critical temperature also occurs near the largest frustration area.
Moreover, it is found that thermal fluctuation stabilizes the N\'{e}el phase even slightly 
above ${J_2/J_1=1/3}$, and thus induces re-entrant behavior. 
This subtle boundary exposes the intricate competitions between the 
spin interactions and thermal fluctuation in spiral spin liquid systems.

\subsection{Specific heat}
\label{sec3C}
As mentioned previously, different classes of spiral spin liquids could exhibit distinct behaviors at low temperatures. 
In addition to the spin correlations, these distinctions can be directly reflected in the thermodynamic quantities. 
To demonstrate, here the specific heats $C_v$ for the spiral spin liquids with $(d_s,d_c)=(1,1)$ and $(1,2)$ respectively, are calculated.

In two-dimensional systems, the spiral spin liquid can persist down to zero temperature.
In the low-temperature regime, fluctuations near the spiral contour govern the physics and the minimum eigenvalue mode vanishes for momenta on the spiral contour.
Therefore, the self-consistent equation \eqref{eq17} can be approximated as 
\begin{equation}
    \int_{\bm{k}}\frac{1}{\omega_{-}(\bm{k})+T^{\alpha}\bar{\lambda}}\sim\frac{2}{3}\beta.
\end{equation}
Here, it is assumed that the saddle-point solution has a power-law temperature dependence as $\lambda'(T)\sim T^{\alpha}\bar{\lambda}$. 
At low temperatures, the integral can be well estimated by considering momenta near the spiral contour.
For a momentum $\bm{q}$ located on the spiral contour, a coordination framework spanned by the normal vector $\hat{\bm{n}}$ and the tangent vector $\hat{\bm{t}}$ can be associated. 
Along the normal direction, the dispersion of the minimum eigenvalue mode can be replaced by $\omega_{-}(\bm{k})=v(\delta p)^2$, where $\delta p = (\bm{k}-\bm{q})\cdot\hat{\bm{n}}$ and $\bm{k}\cdot \hat{\bm{t}}=0$. 
The integral can be taken over $\delta p$ with a proper cutoff and over the contour itself. By modifying the variable $\delta p=T^{\alpha/2} x$, one can scale out the power $\alpha = 2$ for the saddle-point solution $\lambda'(T)$. 

The free energy $F=-T\ln\mathbb{Z}$ in this approximation is 
\begin{equation}
    F \approx -\frac{1}{2}T^2\bar{\lambda}+\frac{1}{2}T\int_{\bm{k}}\ln[(\omega_{-}(\bm{k})+T^2\bar{\lambda})/(2\pi T)].
\end{equation}
To calculate the specific heat $C_v=-T\left(\partial^2 F/\partial T^2\right)_{V,N}$, it is convenient to consider 
\begin{equation}
    \frac{\partial F/T}{\partial T}=-\frac{\bar{\lambda}}{2}-\frac{1}{4\pi T}+\int_{\bm{k}}\frac{T\bar{\lambda}}{\omega_{-}(\bm{k})+T^2\bar{\lambda}}\approx \frac{c_1}{T}+c_2.
\end{equation}
In the last step, the self-consistent equation is used to finish the integral. 
Thus, the free energy has the form $F=c_1 T \ln T +c_2 T^2 + + c_3 T$, and the specific heat is expressed as
\begin{equation}
    C_v = c_1 + c_2 T.
\end{equation}
Here, $c_1$ and $c_2$ are new constants absorbing all coefficients.

For spiral spin liquids in three spatial dimension, the development of spiral orderings in the low-temperature regime has significant effects on the thermodynamic quantities. 
In the absence of the thermal order-by-disorder, the dispersions of fluctuation modes from out-of-plane field $\phi^{\text{o}}$ are $\omega_{\pm}(\bm{k})-\omega_{-}^{\text{min}}$ at the Gaussian level and the lowest one vanishes for any momentum on the spiral contour. 
This inevitably leads to a divergence of the fluctuation amplitude proportional to $T\int_{\bm{k}}1/G_0(\bm{k})$, where $G_0(\bm{k})$ is the propagator of $\phi^{\text{o}}$.
To cure the divergence, one must consider higher-order terms in the expression of fluctuation fields $\phi^{\text{i}}$ and $\phi^{\text{o}}$. 
Their contributions can be formally treated as a self-energy correction $\Sigma(\bm{k})$, giving rise to the Green's function 
\begin{equation}
    G(\bm{k})=\frac{1}{\omega_{\pm}(\bm{k})-\omega_{-}^{\text{min}}+\Sigma(\bm{k})}.
\end{equation}
With the aid of the diagrammatic perturbation theory, $\Sigma(\bm{k})$ can be expressed self-consistently as
\begin{equation}
    \Sigma(\bm{k}) = T\int_{\bm{p}}D(\bm{k},\bm{p})G(\bm{p}),
\end{equation}
where $D(\bm{k},\bm{p})$ is a well-behaved function and vanishes only at finite momenta corresponding to the Goldstone modes. 
Therefore, the dominant contribution of the integral still comes from momenta near the spiral contour. 
The numerator of the integrand is thus replaced by $D(\bm{k},\bm{q})$, where $\bm{q}$ denotes the momentum on the contour. 
Similarly, one can associate the point $\bm{q}$ with a coordination framework spanned by the normal vector $\hat{\bm{n}}$, the tangent vector $\hat{\bm{t}}$, and the original $z$-axis vector $\hat{\bm{z}}$. 
In the $\hat{\bm{n}}$-$\hat{\bm{z}}$ plane, the dispersion of the eigenvalue mode $\omega_{-}(\bm{p})-\omega_{\text{min}}$ is still quadratic. 
One can parameterize momentum $\bm{p}$ in this plane as $\bm{p}\cdot\hat{\bm{n}}=\delta p\cos\theta$ and $\bm{p}\cdot\hat{\bm{z}}=\delta p\sin\theta$, where $\bm{p}\cdot\hat{\bm{t}}=0$ and $\delta p=|\bm{p}-\bm{q}|$.
The dispersion of the minimum eigenvalue mode can then be expressed as $v^2(\delta p)^2$ approximately.
It is further assumed that the self-energy has a temperature dependence with the leading form $\Sigma(\bm{k})=T^{\alpha}\Sigma'(\bm{k})$ due to the thermal splitting. 
The self-consistent equation reduces to 
\begin{equation}
    T^{\alpha}\Sigma'(\bm{k})=T\int_{\bm{p}}\frac{D(\bm{k},\bm{q})}{v^2(\delta p)^2+T^{\alpha}\Sigma'(\bm{q})}.
\end{equation}
The integral is taken on the $\hat{\bm{n}}$-$\hat{\bm{z}}$ plane with respect to $\theta$ and $\delta p$ with a proper cutoff, and over the contour itself.
Modifying the variables as $\delta r=\delta p/T^{\alpha/2}$,
the stiffness $T^{\alpha}$ can be scaled out of the integral and one obtains $\alpha=1$.

The full free energy should include both the $\phi^{\text{i}}$ and $\phi^{\text{o}}$ parts, and be corrected by the self-energy as well. Up to the order of $T$, it takes the form 
\begin{align}
    F= & F_{\phi^{\text{i}}}(\bm{q})+T\int_{\bm{k}}\ln [(\omega_{\pm}(\bm{k})-\omega_{-}^{\text{min}}+T\Sigma'(\bm{k}))/(2\pi T)] \notag \\
    = & c_1 T \ln T + T \int_{\bm{k}}\ln[\omega_{\pm}(\bm{k})-\omega_{-}^{\text{min}}+T\Sigma'(\bm{k})]+c_3 T.
\end{align}
To calculate the specific heat, the following equation is evaluated:
\begin{equation}
    \frac{\partial (F/T)}{\partial T}=  \frac{c_1}{T}+\int_{\bm{k}}\frac{\Sigma'(\bm{k})}{\omega_{\pm}(\bm{k})-\omega_{-}^{\text{min}}+T\Sigma'(\bm{k})}.
\end{equation}
Repeating the same scaling trick, it is found that the temperature dependence of free energy 
$F=c_1T\ln T+c_2 T^2+c_3 T$, and thus the low-temperature specific heat is 
\begin{equation}
    C_v=c_1+c_2 T.
\end{equation}
We have absorbed all coefficients into $c_1$ and $c_2$ and they should not be confused with the previous constants.

% quantum ones. 

\section{Discussion}
\label{secIV}

\subsection{Survey of materials}
\label{sec4A}

A survey of existing materials is presented in which spiral spin liquid physics
can be relevant. The A-site spinels \ce{AB2X4} represent a large family of 
compounds for the diamond-lattice antiferromagnets.      
The magnetism comes from the $A$-site ions that form a diamond lattice.
Among them, \ce{MnSc2S4} and \ce{CoAl2O4} manifest the strongest 
frustration of classical spins and are suggested to support spiral spin liquids 
by early experimental and theoretical works~\cite{PhysRevB.72.174404, 
Suzuki_2007, PhysRevLett.92.116401, Bergman2007}. 
Although later measurements ruled out the possibility of \ce{CoAl2O4} 
as a candidate \cite{PhysRevB.84.094403}, in recent experiments the spiral surface in \ce{MnSc2S4} was directly observed by neutron-scattering 
measurements and proved the existence of a spiral spin liquid regime 
for the spin-$5/2$ \ce{Mn^2+} ions \cite{Gao2016}. 
Instead of an ordering transition via the thermal order-by-disorder, 
a multi-step ordering behavior was observed at low temperatures. 
This reveals the limitation of the $J_1$--$J_2$ Heisenberg model and suggests that
more interactions should be considered, such as the third-nearest-neighbor 
coupling in real materials~\cite{PhysRevB.78.144417, PhysRevB.98.064427}. 
Other spinels like \ce{ARh2O4} (A = Ni, Co, and Cu) and 
\ce{MgCr2O4} have also been synthesized recently~\cite{PhysRevMaterials.2.034404, PhysRevB.96.064413, PhysRevB.96.020412,PhysRevLett.122.097201}. 
For \ce{NiRh2O4} with frustrated spin-$1$ \ce{Ni^{2+}} ions, 
the negative Curie-Weiss temperature and absence of the 
magnetic ordering transition down to 0.1 K are noted. This material has 
spin ${S=1}$ and the quantum effects seem to be strong.
Quantum effects on the spiral manifold were seriously 
considered~\cite{PhysRevB.96.020412,PhysRevLett.120.057201}.  
As the material has a tetragonal symmetry rather than 
cubic symmetry and the \ce{Ni^{2+}} has a partially filled 
$t_{2g}$ shell, the single-ion anisotropy as well as 
the on-site spin-orbit coupling were thus also considered in addition to the superexchange interactions~\cite{PhysRevB.96.020412,PhysRevLett.120.057201,PhysRevB.100.045103,PhysRevB.100.140408}. More experiments 
are needed to resolve the physics in \ce{NiRh2O4}.  
The \ce{Co^{2+}} ion in \ce{CoRh2O4} has ${S=3/2}$, and a large nearest-neighbor exchange was found in \ce{CoRh2O4}, 
so an N\'{e}el order was observed~\cite{PhysRevB.96.064413}.  
The \ce{Cu^{2+}} ion in \ce{CuRh2O4} and \ce{CuAl2O4} 
has a $d^9$ configuration with five electrons on the 
upper $t_{2g}$ shell. The spin-orbit coupling is active. It is not purely
spin physics, and one would necessarily consider the orbital degree of freedom. 
It was noted that this \ce{Cu^{2+}} ion in the tetrahedral environment is similar 
to the well-known \ce{Ir^{4+}} ion in the octahedral environment and gives
a ${J_{\text{eff}} = 1/2}$ local moment~\cite{PhysRevB.100.045103,PhysRevB.98.201106,XPY2020}. 
The model for these \ce{Cu^{2+}} moments in \ce{CuRh2O4}
would not be a simple $J_1$--$J_2$ model, so spiral spin liquid physics
may not be directly relevant. This system would be better described 
by an anisotropic Heisenberg-compass model on the diamond lattice. 
This will be analyzed in future works. 
Recently, a rare-earth compound \ce{LiYbO2} was proposed to realize
a diamond lattice~\cite{bordelon2020frustrated}. Its theoretical understanding
was based on the effective spin-1/2 $J_1$--$J_2$ Heisenberg model with the proximity to 
spiral spin liquid physics and the spiral manifold despite a large spin-orbit
coupling for the \ce{Yb^{3+}} ions. The possible success in \ce{LiYbO2}
may point to the possibility of realizing a spiral spin liquid in
other Yb-based systems, such as the Yb-based honeycomb-lattice magnets.  

The 6H-B-\ce{Ba3NiSb2O9} compound with a spin ${S=1}$ local moment
has the same structure of the AB-stacked multilayered triangular lattice model studied herein,
and this material was proposed to be a spin liquid 
candidate~\cite{PhysRevLett.107.197204}. 
It has been suggested that the quantum effect in addition to 
spiral contour physics plays an important role 
in the spin-liquid phenomenology~\cite{PhysRevLett.109.016402}. 
This may inspire the study of isostructural materials with ${d_s=1}$ spiral manifolds 
in three dimensions. The compound \ce{MgCr2O4} was shown to 
have line degeneracies in reciprocal space, and the spiral spin liquid 
was proposed to be relevant~\cite{PhysRevLett.122.097201}. 
This again corresponds to the ${d_s =1}$ case.

The relevant materials in two dimensions can be \ce{Bi3Mn4O12(NO3)} 
and \ce{Ca10Cr7O28}. \ce{Bi3Mn4O12(NO3)} is a bilayer honeycomb-lattice magnet with $S=3/2$ spins and exhibits a spin-liquid-like behavior 
down to 0.4 K~\cite{doi:10.1143/JPSJ.79.114705}. 
Theoretically, this observed behavior falls into the situation of massive ring-like degeneracy and an order-by-disorder mechanism in a $J_1$--$J_2$ 
Heisenberg model~\cite{PhysRevLett.105.187201,doi:10.1143/JPSJ.79.114705}. 
Recently, a novel ripple state under the external magnetic field 
has been reported and can be explained by the infinite ring-like 
degeneracy~\cite{PhysRevLett.123.057202}. \ce{Ca10Cr7O28} 
is another frustration magnets in which \ce{Cr^{5+}} ions 
form a Kagom\'{e} bilayer structure. 
As the interaction between the \ce{Cr^{5+}} spin-1/2 moment inside 
the triangular unit is dominantly ferromagnetic, one can treat the
triangular unit as a total spin ${S=3/2}$. 
The physics of this bilayer system can then be modeled 
as a semiclassical Heisenberg Hamiltonian 
on a honeycomb lattice with effective spin ${S=3/2}$ moments
and the intralayer and interlayer couplings are antiferromagnetic 
and ferromagnetic, respectively~\cite{PhysRevB.97.115102}.
A ring-like pattern in the neutron-scattering signal and ordering transition at low temperature are also suggested for these materials~\cite{pohle2017spin, PhysRevB.97.115102}. These studies suggest a particular relevance to the spiral spin liquid.
However, there is still a lack of experimental results to reach a conclusion at this stage.

\subsection{Open theoretical questions}
\label{sec4B}

Several open theoretical questions about the physics
related to the spiral spin liquids are discussed here. In our calculation, the spin is treated classically. 
However, in reality, the spins always have quantum effects, especially for small spins. 
It was well known that the quantum fluctuations could break the spiral degeneracy
for the ground state via the mechanism of the quantum order by disorder,
and specific spiral states could be selected. The ordered states via the quantum order 
by disorder can sometimes be different from the thermal order by disorder. 
The thermal order by disorder is an entropic effect and requires a finite temperature,
while the quantum order by order results from the quantum zero-point energy 
and works well for zero and low temperatures. 
This seems to suggest that there could be two thermal transitions as one cools 
the system from the high-temperature paramagnetic phase. 
The system first enters the ordered state via the thermal order by disorder
and then develops another ordered phase due to the quantum order by disorder at
a lower temperature. This is possible. 
Nevertheless, the separation of quantum order by disorder and 
thermal order by disorder is an artificial procedure, while nature blends these
two effects together in the system. The blending of these effects may even favor 
another ordered state that differs from that favored by either. 
These issues may be resolved in future theoretical works.

Another important question concerns the internal spin space.  All the models
considered in this work have a global rotation symmetry with Heisenberg interactions 
and spins. The question is whether the spiral spin liquid regime
can persist for XY spins or even for Ising spins. For XY spins with a U(1) symmetry,
it is still possible to construct spiral states with two components of the 
XY spins. Thus, it is natural to expect that the spiral spin liquid exists for the XY
or Heisenberg spins with an easy-plane anisotropy. 
Moreover, the senario for sprial spin liquids with XY spins resembles the bosonic models with frustrated hoppings, in which bosons could condense in a closed line instead of discrete points. 
Such phenomena have been discussed in various systems including the frustrated systems \cite{PhysRevLett.114.037203}, high $T_c$ superconductors \cite{PhysRevB.99.104507}, and driven-dissipative systems \cite{PhysRevLett.125.115301}.
For Ising spins, it is not obvious if such a spiral spin liquid 
could survive as there is no way that one can construct 
spiral spin states with Ising spins. In contrast, the self-consistent
Gaussian approximation that ignores the individual spin-magnitude constraint
works rather well for the spiral spin liquid, and 
was known to work reasonably for certain Ising systems. 
Taken together, it is uncertain whether 
Ising spins could support the spiral spin liquid regime;
this question will be addressed in a forthcoming paper. 

Finally, the interplay of the quantum effect and degenerate
 spiral manifold may require a more systematic study in the future. 
Only in a few previous works, including 
Refs.~\onlinecite{PhysRevLett.120.057201,PhysRevB.96.020412,PhysRevB.100.045103,PhysRevLett.101.047201} 
in the diamond lattice antiferromagnet and 
Refs.~\onlinecite{PhysRevLett.109.016402,PhysRevB.81.214419,Niggemann_2019} in 
the honeycomb lattice, and the AB-stacking multilayered triangular 
lattice, have explorations along this line been conducted.  
In fact, many of the spin moments in the existing materials already 
involve significant quantum fluctuations and could provide a good
experimental platform for studying these physics.

\subsection{Summary}
\label{sec4C}

In summary, the spiral spin liquids based on the dimension $d_s$ 
and codimension $d_c$ of the degenerate spiral manifold are classified. For each class, 
a minimal model based on the $J_1$--$J_2$ Heisenberg Hamiltonian is constructed. 
These constructions are not unique, but essentially capture the physics of spiral 
spin liquids. Systematic studies are performed on the square, honeycomb, and 
AB-stacked multilayered triangular lattices. These lattices, together with the well-studied 
diamond lattice, cover all three classes in the classification presented in Table~\ref{tab:summary}.
The present model construction takes the bipartite lattice as a starting point. 
The non-Bravais lattices, like the honeycomb and diamond lattices, have 
been studied and incorporated into this work and the classification.
Many Bravais lattices, such as the square lattice in two dimensions, 
simple cubic, face-centered-cubic, and body-centered cubic lattices in three 
dimensions, can be treated as bipartite lattices as well once their 
adjacency matrices are carefully dealt with. The penalty of this 
straightforward extension is the requirement of further and more 
complicated interactions that may be rather unnatural in real materials. 
It is likely that some materials may be located near the unnatural parameter 
regime, and the additional interactions can be treated as a perturbation 
to break the spiral manifold. Despite this, the current classification and theoretical 
results depend only on the dimensions and can be immediately applied 
once the spiral manifold emerges.

\begin{acknowledgments}
We thank Dr Fei-Ye Li and Dr Xiao-Tian Zhang for useful discussions. 
This work is supported by the Ministry of Science and Technology of China 
(Grant Nos.~2018YFE0103200, 2016YFA0300500, and 2016YFA0301001), 
the Shanghai Municipal Science and Technology Major Project (Grant 
No. 2019SHZDZX04), and the Research Grants Council of Hong Kong 
with General Research Fund (Grant Nos. 17303819 and 17306520). 
\end{acknowledgments}

\appendix

\section{Lattice geometry and crystal momenta}
\label{appI}
The square lattice is regarded as a bipartite lattice with two sublattices as shown in Fig.~\ref{fig:lattices}(a) and one of the unit cells is chosen as a reference point with sublattice positions $(0,0)$ and $(1,0)$. The primitive lattice vectors are
\begin{equation}
    \bm{a}_1=(1,1),\  \bm{a}_2=(-1,1).
\end{equation}
In reciprocal space, two reciprocal vectors are defined by
\begin{equation}
    \bm{b}_1=(\pi,\pi),\  \bm{b}_2=(-\pi,\pi).
\end{equation}
The crystal momenta shown in Fig.~\ref{fig:square}(a) are
\begin{equation}
    X=\left(\frac{\pi}{2},\frac{\pi}{2}\right),\  M=\left(\pi,0\right).
\end{equation}

For the honeycomb lattice, a unit cell containing sublattices at positions $(0,0)$ and $(1/\sqrt{3},0)$ is chosen as a reference point. The primitive lattice vectors are
\begin{equation}
    \bm{a}_1=\left(\frac{\sqrt{3}}{2},\frac{1}{2}\right),\  \bm{a}_2=\left(\frac{\sqrt{3}}{2},-\frac{1}{2}\right),
\end{equation}
and the reciprocal vectors are defined by
\begin{equation}
    \bm{b}_1=2\pi\left(\frac{1}{\sqrt{3}},1\right),\  \bm{b}_2=2\pi\left(\frac{1}{\sqrt{3}},-1\right).
\end{equation}
The crystal momenta shown in Fig.~\ref{fig:honeycomb}(b) are
\begin{equation}
    M=2\pi\left(\frac{1}{\sqrt{3}},0\right),\  K=2\pi\left(\frac{1}{\sqrt{3}},\frac{1}{3}\right).
\end{equation}

The AB-stacked multilayered triangular lattice is a hexagonal system. Two sites connected by interlayer interaction $J_1$ are chosen as an unit cell and their positions set to be $(0,0,0)$ and $(1/2,\sqrt{3}/6,c/2)$, where $c$ is a positive constant. The primitive vectors then reads
\begin{equation}
    \bm{a}_1=(1,0,0),\  \bm{a}_2=\left(\frac{1}{2},\frac{\sqrt{3}}{2},0\right),\  \bm{a}_3=\left(0,0,2c\right).
\end{equation}
The reciprocal vectors are 
\begin{equation}
    \bm{b}_1=\left(2\pi,-\frac{2\pi}{\sqrt{3}},0\right),\  \bm{b}_2=\left(0,\frac{4\pi}{\sqrt{3}},0\right),\  \bm{b}_3=\left(0,0,\frac{\pi}{c}\right).
\end{equation}
The crystal momenta shown in Fig.~\ref{fig:ABcontour} are 
\begin{equation}
    M=2\pi\left(0,\frac{1}{\sqrt{3}},0\right),\  K=2\pi\left(\frac{1}{3},\frac{1}{\sqrt{3}},0\right).
\end{equation}

\bibliography{SSLRef.bib}

\end{document}